\documentclass[twocolumn,trackchanges]{aastex701}

\usepackage{hyperref}
\usepackage{comment}
\usepackage{gensymb}
\usepackage{tablefootnote} 

\hypersetup{
    colorlinks=true,
    linkcolor=magenta,
    filecolor=magenta,
    citecolor=blue,
    urlcolor=magenta,
    pdfpagemode=FullScreen,
    }

\received{by ApJ, January 5, 2026}
\revised{January 9, 2026}
\accepted{January 17, 2026}

\begin{document}

\title{The First Quantitative Study of Tail Regrowth of CME-Driven Disconnection in Comet C/2023 P1 Nishimura}

\author[0000-0002-3089-3431]{Shaheda Begum Shaik}
\affiliation{George Mason University, Fairfax, VA 22030, USA}
\affiliation{U.S. Naval Research Laboratory, Washington, DC 20375, USA}
\email[show]{sshaik7@gmu.edu}

\author[0000-0001-8480-947X]{Guillermo Stenborg}
\affiliation{The Johns Hopkins University Applied Physics Laboratory, Laurel, MD 20723, USA}
\email[]{guillermo.stenborg@jhuapl.edu}

\author[0000-0003-1377-6353]{Phillip Hess}
\affiliation{U.S. Naval Research Laboratory, Washington, DC 20375, USA}
\email[]{phillip.n.hess2.civ@us.navy.mil}

\author[0000-0002-8164-5948]{Angelos Vourlidas}
\affiliation{The Johns Hopkins University Applied Physics Laboratory, Laurel, MD 20723, USA}
\email[]{angelos.vourlidas@jhuapl.edu}

\author[0000-0002-8692-6925]{Karl Battams}
\affiliation{U.S. Naval Research Laboratory, Washington, DC 20375, USA}
\email[]{karl.battams.civ@us.navy.mil}

\author[0000-0002-3253-4205]{Robin Colaninno}
\affiliation{U.S. Naval Research Laboratory, Washington, DC 20375, USA}
\email[]{robin.c.colaninno.civ@us.navy.mil}

\begin{abstract}
Comet C/2023 P1 (Nishimura) was observed by the Solar Orbiter Heliospheric Imager (SoloHI), onboard the Solar Orbiter spacecraft, from 2023 September 1 to 14. During this period, the ion tail of the comet exhibited continual fluctuations and four tail disconnection events (TDEs), each coinciding with the passage of a coronal mass ejection (CME). In this work, we report on the ion tail dynamics of the best observed TDE, which occurred on September 11. The SoloHI white-light images reveal an abrupt bending, subsequent kinks, and severing of a downstream portion of the pre-existing ion tail. The onset of disconnection occurred $\sim$6.5 hours after the projected passage of the CME leading edge in the images, consistent with a CME flank encounter. After the disconnection, the ion tail reformed within $\sim$24 hours, with a regrowth rate of $\sim$86$\pm7~\mathrm{km\,s^{-1}}$, indicating the rate at which newly ionized material forms along the magnetic field draped around the comet's coma. After the TDE, the detached tail drifted anti-sunward at an estimated speed of $\sim$295\(\pm20~\mathrm{km\,s^{-1}}\), comparable to the local CME flank's speed, suggesting that the severed plasma was most likely carried away from the comet by the CME. This study provides the first direct, quantitative characterization of comet-CME interactions and the subsequent regrowth phase of a cometary TDE. These measurements were achievable by SoloHI's unique inner-heliospheric coverage, thanks to a combination of high photometric sensitivity, short exposure times, and a wide field of view that preserves the fine-scale tail dynamics. 
\end{abstract}


\section{Introduction}\label{sec_intro}
The ion (plasma) tail of a comet acts as a natural probe of heliospheric conditions, as its orientation and brightness respond to variations in ambient solar wind velocity, density, and interplanetary magnetic field (IMF) topology \citep{Biermann_1951, Alfven_1957}. Observations of cometary activity shed light on the plasma processes governing outgassing, ionization, and tail formation, which are influenced by the solar environment \citep[for example,][]{Ip_2004, Broiles_2015, Behar_2018, Bemporad_2023}. As the comet approaches the Sun, the sublimating gases from the comet nucleus form a coma, within which neutral molecules become ionized \citep{Povich_2003}. These ions are highly responsive to the solar wind and the IMF, leading to an antisunward ion tail aligned with the local IMF direction. In contrast, the dust tail consists of neutral dust grains, which are sorted by solar gravity and radiation pressure, producing a broad tail mostly aligned with the curved comet trajectory \citep[][and references within]{Whipple_1950, Jones_2017}.

During the comet's passage inside of 1 astronomical unit (au) from the Sun, the ion tail exhibits a dynamic nature, characterized by oscillations, instabilities, and significant amplitude kinks and knots \citep{Nistico_2018, Wellbrock_2025}. The interplanetary magnetic field carried by the solar wind and draped around the comet’s ionized coma can reconnect with solar magnetic structures, leading to the detachment of the tail, creating a tail disconnection event \citep[TDE;][]{Barnard_1899, Barnard_1920, Voelzke_2005}. The principal heliospheric drivers for these TDEs are reported as the crossing of interplanetary sector boundaries such as the heliospheric current sheet \citep[HCS;][]{Niedner_1978, Yi_1996, Konz_2004, Jia_2007}, encounters with stream interaction regions \citep[SIRs;][]{Wegmann_2000}, and interaction with the magnetic structure of coronal mass ejections \citep[CMEs;][]{Jones_Brandt_2004, Vourlidas_2007, Kuchar_2008, Jia_2009}. 

After the disconnection, the tail rapidly reforms as the cometary material continues to ionize. Although many TDEs have been reported since the 1970s, quantitative constraints on the timescale for tail reestablishment and its dependence on local plasma parameters have seldom been achieved, owing to limitations such as time cadence, restricted fields of view (FOVs), and the difficulty in resolving the finer details and dynamics of the tail in many comet observations. As a result, reformation has been described as instantaneous \citep{Voelzke_2005} on observing timescales, without the benefit of resolved measurements. The most dramatic comet dynamics emerge as comets migrate into the inner heliosphere, where heating and rapid IMF variability are pronounced. However, Earth-based observations in this domain are often challenging because small solar elongation angles suppress comet visibility, making heliospheric imaging instruments the primary means of continuously capturing these dynamic detections.  

The Solar Orbiter Heliospheric Imager \citep[SoloHI;][]{Howard_2020} is a visible light imager onboard the Solar Orbiter spacecraft \citep{Muller_2020, Garcia_2021} that can resolve finer features in the heliosphere than comparable instruments at $1$~au \citep{Hess_2023}. SoloHI's improved spatial and temporal resolution, along with a $40^{\circ}$ wide FOV, enable detailed tracking of fast morphological changes over considerable distances in the inner heliosphere. 

In this study, we analyze the tail dynamics of comet C/2023~P1 \citep[Nishimura;][]{Maxim_2024} utilizing unique comet observations from SoloHI. We focus on interactions between the comet, CMEs, and the solar wind over an extended observing period spanning from 2023 September 1 to 14, quantifying the evolution of the ion tail and its response to transient and ambient heliospheric conditions. During this period, four TDEs were observed on September 3, 5, 8, and 11, each coinciding with the crossing of a CME. During this period, SoloHI offered a unique opportunity to capture comet-solar wind-CME interactions, allowing the tracking of fine details and resolving the plasma tail dynamics from the nucleus of the comet. All four of these disconnections coincide with CME observations, indicating that CME-driven magnetic reconnection as the dominant trigger for these events, rather than heliospheric current sheet or stream interaction region crossings. 

To better constrain the propagation direction and velocity of the CMEs observed in SoloHI, we also utilize white-light images from the LASCO/C2 and C3 \citep[][]{Brueckner_1995} coronagraphs on board the Solar Heliospheric Observatory \citep[SOHO;][]{Domingo1995} and the COR2 \citep{Howard_2008} coronagraph onboard the Solar TErrestrial RElations Observatory-A \citep[STEREO-A;][]{Kaiser2008}. These instruments, located near 1 au and positioned close together in longitude during the comet observation (Figure~\ref{Figure1}), provide crucial context for interpreting the CMEs that were later observed by SoloHI at larger distances in the heliosphere.

Comet C/2023 P1 was also imaged in white-light by the Wide Field Imager for Solar Probe (WISPR; \citealt{Vourlidas2016}), on board Parker Solar Probe \citep[PSP;][]{Fox2016, Raouafi_2023} between 2023 September 27 -- 28, and by the STEREO-A/HI-1 imager \citep[][]{Eyles_2009}, between 2023 September 17 -- 30. A study of comet dynamics based on these observations and a three-dimensional viewing geometry analysis is planned for a forthcoming article.

We present our findings on the best observed TDE, which occurred on 2023 September 11, to quantify the dynamics of disconnection, investigate its correlation with the CME crossing, monitor the onset, and examine the regrowth of the post-disconnection tail. The manuscript is organized as follows: Section~\ref{sec:obs} describes the observations and details of the comet ephemeris, Section~\ref{sec:discussion} presents the results and discussion, and finally, Section~\ref{sec:conclusion} summarizes our findings.

\section{Observations}\label{sec:obs}
\subsection{SoloHI Observations of Comet C/2023 P1}\label{sec:Comet_obs}

\begin{figure*}
\begin{center}
\includegraphics[width=18cm]{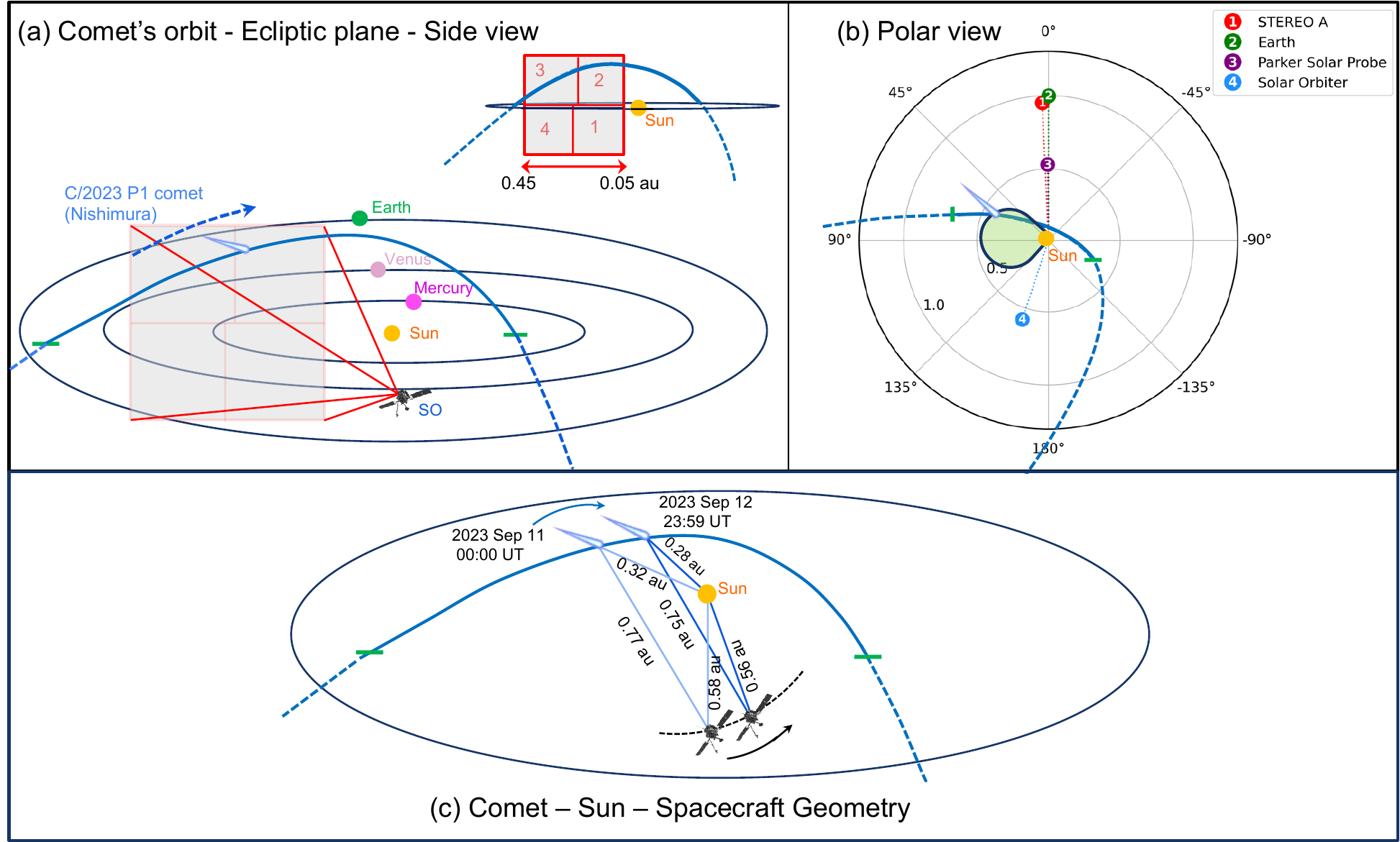}
\end{center}
\caption{Trajectory of C/2023 P1 comet, with the respective positions of Solar Orbiter, the nearby planets, and other spacecraft. (a) Side view of the comet location relative to the ecliptic plane on 2023 September 11. The blue dashed arrow indicates the direction of the comet trajectory. The path of the comet above the ecliptic plane is shown in solid blue, and below in dashed blue. An inset shows the SoloHI FOV (gray shaded region enclosed by the red outline) on that day. (b) The ecliptic plane and the comet's path as viewed from above the solar north pole. The green-shaded region shows the extent of the CME associated with the TDE under study and illustrates the geometry of CME-comet interaction. The polar map in the background layer is produced using the Solar-MACH \citep{gieseler_solar-mach_2023}. (c) Comet-Sun-Solar Orbiter geometry, the comet and Orbiter trajectory during the analysis period of 2023 September 11, 00:00 UT to 12, 23:59 UT. The orbital ephemerides are obtained from the JPL Horizons page. This schematic renders the simplified CME-comet geometry and is not drawn to scale.}
\label{Figure1}
\end{figure*}

The Solar Orbiter spacecraft is in an elliptical orbit around the Sun, with a perihelion of 0.28 au. Between 2023 September 01 and 14, the spacecraft's orbital distance dropped from 0.69 to 0.54 au. The onboard imager, SoloHI, observes the solar corona and inner heliosphere in visible light (450-850 nm), with a FOV of $\sim$$5$ to $45^{\circ}$ from the Sun's center, on the anti-ram side of the spacecraft. The SoloHI images record the photospheric light scattered by electrons \citep[via Thomson scattering, i.e., the K-corona;][]{Billings1966}, and by the interplanetary dust particles orbiting the Sun \citep[i.e., the F-corona;][]{Kimura_1998, Lamy_2022}. Additionally, interplanetary bodies and the star field complete the scene inside the FOV.

The SoloHI images are acquired with an array of four active-pixel sensors (APS) tiles (numbered 1 to 4 counterclockwise as shown in Figure~\ref{Figure1}a). The tiles are exposed and read out independently, and can be combined to form a single image on the ground. This configuration allows individual tiles to have their exposure times and cadence optimized. These images\footnote{All SoloHI data are publicly accessible at the SoloHI homepage (\url{https://solohi.nrl.navy.mil/so_data})} undergo a calibration process to eliminate detector bias, normalize the signal for exposure time, and convert the detected intensities into units of mean solar brightness (MSB; \citealt{Hess_2023}).

To remove the F-corona component and hence reveal the much fainter K-corona signal, we further process the calibrated images using the `LW' technique \citep[Appendix A of][]{Howard_2022}, which was initially developed for PSP/WISPR, a similar white-light heliospheric imager. This technique leverages the time domain to model the background signal, thereby eliminating the influence of the considerable pseudo-stationary component of the signal, which is the largest contributor to the intensity of the images \citep{Stenborg_2025}. 

\begin{table*}[!ht]
\centering
\caption{Disconnection events during the comet observing period}
\begin{tabular}{|m{1.9 cm} | m{3.1cm}| m{3.1cm} |  m{2.5cm} | m{2.6cm} |} 
\hline 
\multicolumn{2}{|c}{SoloHI Observations} & \multicolumn{3}{|c|}{LASCO Observations} \\
\hline
Disconnection events & CME-comet crossing Date and time & CME eruption Date and time & Source location & Angular width (degrees)\\ 
\hline \hline
TDE 1 & 2023 Sep 04, 01:43 UT&2023 Sep 03, 09:12 UT&Western limb&249 (Partial Halo)\\
\hline
TDE 2 & 2023 Sep 06, 04:55 UT&2023 Sep 05, 14:00 UT&Western limb&360 (Halo)\\
\hline
TDE 3 & 2023 Sep 09, 03:23 UT&2023 Sep 08, 11:36 UT&Western limb&360 (Halo)\\
\hline
\textbf{TDE 4}&2023 Sep 11, 22:35 UT&2023 Sep 10, 16:48 UT&Western limb&126 (partial halo)\\
\hline
\end{tabular}
\label{table1}
\hfill
\end{table*}

\subsection{Comet Ephemeris}\label{sec:ephemeris}
C/2023 P1 is a long-period comet, discovered by Hideo Nishimura on 2023 August 12 in Kakegawa, Japan.  Its perihelion occurred on 2023 September 17 at a heliocentric distance of $0.225$~au. Aphelion is at $115.32$~au and the orbital period is about 439 years. The comet's orbit is highly eccentric, $e=0.996$. At its perihelion, the comet's apparent visual magnitude was $2.5$, while its total absolute magnitude was $12.2$. 

Figure~\ref{Figure1} illustrates the orbital trajectory of the comet and the observing geometry of SoloHI. The blue curve delineates the comet's path (above the ecliptic in solid and below in dashed) and the location of the comet on 2023 September 11. In panel (a), the gray shaded region, bounded by the red outline, represents the SoloHI FOV of the four APS tiles. The inset cartoon shows the SoloHI FOV from the Sun's center and extending across the ecliptic plane, with the projected orbital path of the comet depicted in blue. 
Panel (b) is the view from above the solar north, on 2023 September 11, showing the comet path and the respective locations of Earth, Solar Orbiter, STEREO-A, and PSP spacecraft. LASCO/SOHO is located at the Lagrange point L1 near Earth. During this observing period, Solar Orbiter was almost in solar conjunction with the Earth, as Earth and Solar Orbiter were separated by nearly 160$^{\circ}$. The green shaded region indicates the approximate longitudinal extent of the CME propagation direction associated with the TDE under study. Panel (c) shows the evolution of the comet-Sun-Solar Orbiter geometry from JPL Horizons\footnote{\url{https://ssd.jpl.nasa.gov/tools/sbdb_lookup.html\#/?des=2023\%20P1&view=VOP}} during the analysis period of the TDE under study (2023 September 11, 00:00 UT to September 12, 23:59 UT).
 
The radial extent of SoloHI FOV depends on the spacecraft’s heliocentric distance. However, since the spacecraft moved only from 0.58 to 0.56 au during the time period of the TDE, the linear extent of FOV remained nearly constant. As illustrated in Figure~\ref{Figure1}c, the distance between the comet and the spacecraft changed from 0.77 to 0.75 au. Consequently, the linear extent of a SoloHI pixel decreases from $\sim$$41.6~\,\mathrm{Mm}$/pixel to $\sim$$40.5~\,\mathrm{Mm}$/pixel, resulting in a change of $\sim$$2.6\%$. This variation can be considered negligible as it falls well below the pixel-level measurement uncertainties. Using these values, the approximate linear extent of the tile 2 FOV, in which the main analysis is conducted, is from about 10 to 55~$R_\odot$.

\section{Results and Discussion}\label{sec:discussion}
\subsection{Disconnection Events and CME Dynamics}\label{sec:CMEkin}

\begin{figure*}[ht!]
\begin{center}
    \includegraphics[width=18cm]{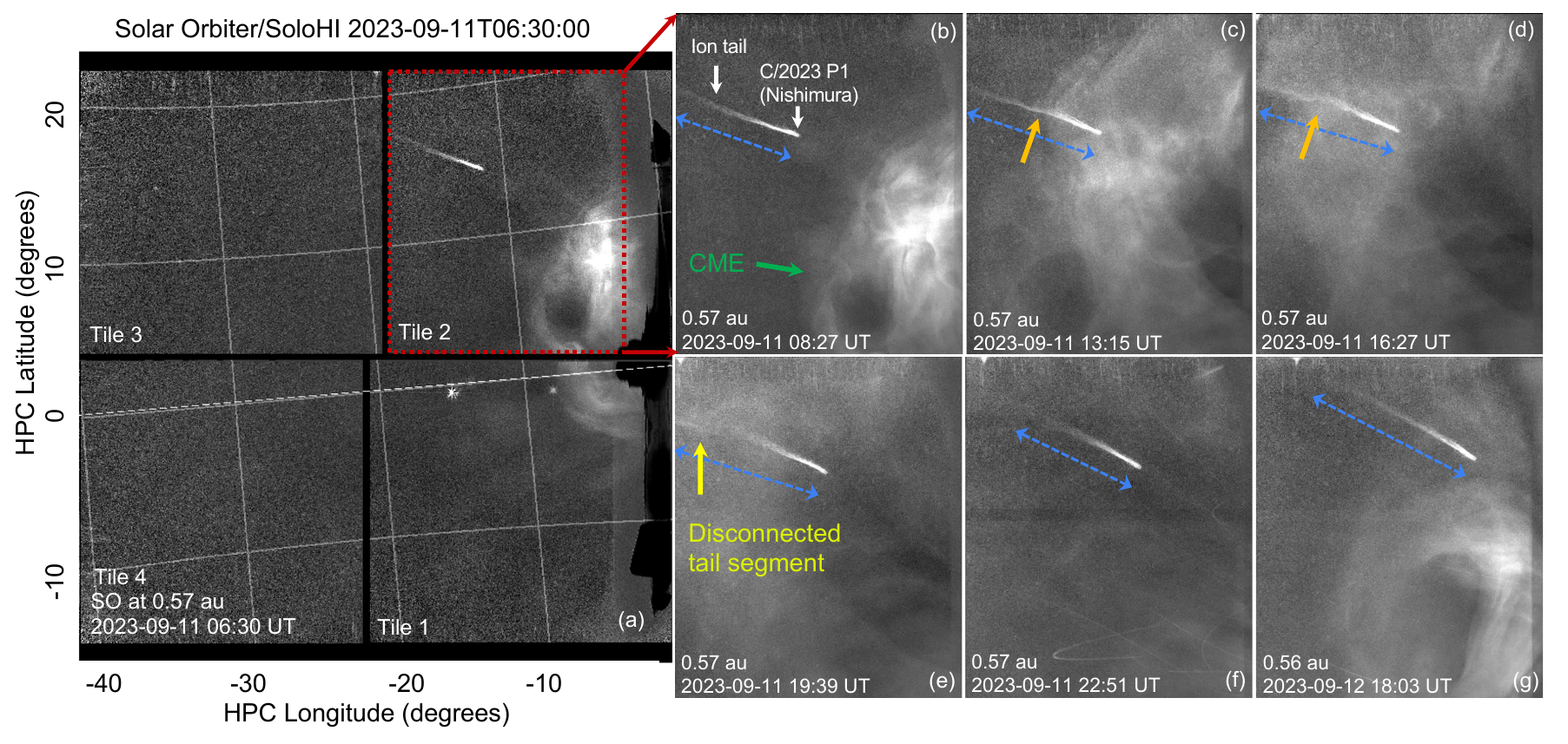}
\caption{SoloHI observations of the C/2023~P1 comet-CME interaction on 2023 September 11. (a) Snapshot of the full SoloHI mosaic FOV, where the Sun is to the right of the image. The grid, delineated by continuous lines, depicts the longitude and latitude in the HPC coordinate system, and the dashed line indicates the ecliptic plane. (b--g) Time sequence of selected image frames of the subfield marked in a red box of panel (a) showcasing the comet-CME interaction. The orange arrows highlight the kink in the tail, and the yellow arrow indicates the detached tail. After disconnection, tail regrowth is shown in the last two snapshots of the sequence, as indicated by the tail length and intensity, highlighted by the blue dashed arrows. A movie illustrating the overall evolution of the comet, its interactions with CMEs and TDEs, is available in the online version of the article. The movie first presents the wide-field evolution in the mosaic view of tiles 3 and 2, followed by a zoomed-in view of the selected subfield.}
\label{Figure2}
\end{center}
\end{figure*}

Comet C/2023 P1 was first detected by SoloHI on 2023 September 1 at 08:50 UT, and remained in the instrument's FOV until September 14 at about 04 UT. The evolution of the comet and its plasma tail can be fully appreciated in the SoloHI mosaic and subfield movie available in the online version of Figure~\ref{Figure2}. 

The mosaic view in the movie shows the overall evolution of the comet within the SoloHI FOV and multiple disconnection events in its tail. The observed tail is interpreted as an ion tail based on its narrow, highly collimated morphology, persistent anti-sunward orientation, and abrupt disconnection followed by rapid regrowth. Such behavior is characteristic of ion tails governed by the interplanetary magnetic field during interactions with transient solar wind structures, CMEs, and is not exhibited by dust tails, which evolve smoothly under solar radiation pressure \citep[e.g.,][]{Niedner_1981, Brandt_1992}.

The occurrence times and correlated CMEs \citep[as listed in the LASCO CME catalog\footnote{\url{https://cdaw.gsfc.nasa.gov/CME_list}};][]{Yashiro_2004} for each TDE are summarized in Table~\ref{table1}. Among the four disconnection events, TDE 4 stands out as the best observed, as it is the only TDE captured in tile 2, which has a higher cadence and shorter exposure time (compared to tile 3), and is therefore the focus of this study. The accuracy of the measurements benefits from tile 2's relatively short exposure times, which minimize motion blur, noise, and preserve tail morphology even during rapid evolution. During our analysis period, SoloHI tile 2 exposure times of 55 to 65 s correspond to $<$1 pixel of motion for a given feature moving at \(250\,\mathrm{km}\,\mathrm{s}^{-1}\) speed.

The comet entered the tile 2 FOV on 2023 September 9 at 08:27 UT, and TDE 4 occurred during the crossing of a CME on September 11. In panel (a) of Figure~\ref{Figure2}, we present a SoloHI mosaic image that encompasses the four APS tiles. Since the images were acquired at different times for the four tiles, the reported time corresponds to the average observation time of the individual tile. The grid delineates the longitude and latitude in a Helioprojective cartesian \cite[HPC;][]{Thompson_2006} coordinate system, which is observer-centered with the Sun at the 0$^\circ$ longitude and 0$^\circ$ latitude. The red rectangle in tile 2 highlights the region of interest where the comet and tail can be observed in detail. Panels (b) through (g) of the figure illustrate six instances of the comet’s evolution within this subfield.

The CME associated with TDE 4 was detected in the LASCO/C2 and SECCHI-A/COR2 coronagraph FOVs on 2023 September 10, with its first appearance in the C2 coronagraph at $\sim$16:48 UT. The source region of the CME on the Sun was located near the solar west limb, relative to the Earth, and the angular width of the CME is denoted by the green region in Figure~\ref{Figure1}b.

The LASCO CME catalog reports a plane of sky (POS) speed of $\sim$\(406\,\mathrm{km}\,\mathrm{s}^{-1}\) for this CME based on a linear fit of height-time measurements up to $\sim$30 $R_\odot$. However, this value reflects only the projection of the CME motion onto LASCO’s viewing plane and therefore does not represent the true physical speed of the CME. Recovering the actual deprojected propagation speed requires determining the three-dimensional (3D) geometry of the CME and measuring its height along the nose direction. Moreover, with POS measurements alone, it is challenging to unambiguously identify the 3D CME-comet interaction conditions directly in the image plane, as they do not indicate which region of the CME's 3D curved surface (nose or flank) encounters the comet.

To address this geometric ambiguity, we reconstruct the 3D CME geometry utilizing the Graduated Cylindrical Shell model \citep[GCS;][]{Thernisien_2011}. This model is a forward-modeling tool that allows a user to adjust a set of six free parameters to reproduce the visible CME flux rope structure in at least one image and is widely used for CME reconstructions \citep{Kay_2024}. While the method can work on a single image from a single viewpoint, the uncertainties are significantly reduced if multiple images, as close in time as possible and from different vantage points, are utilized together \citep{Nikou_2025}. We perform the reconstruction over the interval 18:30 UT on 10 September to 22:00 UT on 11 September using observations from LASCO/C2, C3, SECCHI-A/COR2, and SoloHI. As the images are not co-temporal across the entire CME propagation interval, we use overlapping instrument pairs at their respective times of coverage. In particular, the LASCO/C3 and SoloHI overlap during the later phase provides the strongest constraints on the CME geometry and kinematics at the heights and times relevant to the TDE. This ensures that the derived propagation direction and speed are anchored to the observations that best correspond to the CME conditions at the event location.

The heliocentric heights of the CME leading edge obtained from the GCS reconstruction increase from $\sim$4 to 80 $R_\odot$ ($\sim$0.02 to 0.37 au) during the fitting interval. From a linear fit of these heights, we derive a deprojected CME nose speed of $\sim$\(560\pm15\,\mathrm{km}\,\mathrm{s}^{-1}\). The error estimate on speed and the other errors reported in the manuscript are from the standard error of 1$\sigma$ uncertainty of the fitted slope. 

The best-fit parameters from GCS provide the CME propagation direction at (longitude, latitude) $\approx$ (110$^\circ$, 25$^\circ$) in Stonyhurst coordinates, with a CME half-angle of 52$^\circ$. When combined with the aspect ratio ($\kappa$ = 0.36), this half-angle produces the GCS angular extent of $\sim$$146^\circ$ in the longitudinal direction. Additionally, the obtained tilt angle of the flux rope of $15^{\circ}$ and the comet's trajectory indicate that the comet encounters and traverses through the broad flank region of the CME at a heliocentric distance of $\sim$64 $R_\odot$ as it propagates across the inner heliosphere. In the SoloHI viewpoint, this flank corresponds to the portion of the CME on the far side from the instrument, as illustrated in Figure~\ref{Figure1}b. As white-light imaging observes the heliosphere in integrated Thomson-scattered light along extended lines of sight (LOS), structures located at different depths can appear superimposed in the same region of the image plane. As a result, the CME leading edge can be seen passing over the comet in the projection before the actual physical encounter at the $\sim$64 $R_\odot$ flank location. These LOS projection effects explain the co-spatial appearance of CME and comet in the image sequence, well in advance of the actual CME–comet interaction.

Using the flux rope geometry of the flank region \citep[as shown in \cite{Thernisien_2011}, the Appendix section of][]{Hess_2015} and the angular separation between the CME nose and the comet (evolving from $\sim50-60^\circ$), we estimate the local speed of the CME at the comet's position in the flank region as $\sim$\(380\pm20\,\mathrm{km}\,\mathrm{s}^{-1}\). This value is lower than the nose speed and is consistent with the reduced expansion speed expected at large angular distances from the CME nose under self-similar evolution.

\subsection{Dynamics of the Ion Tail and Disconnection Event 4}

As seen in panels (b and c) of Figure~\ref{Figure2} and in the subfield view in the movie, tiny, intermittent fluctuations in the ion tail are ubiquitous during the comet's trajectory through the FOV, indicating its interaction with the ambient medium, before the CME reaches the comet. At around 13:15 UT, the tail shows deflection and a kink (orange arrows in panels c and d). The origin of this kink is not definitively established, but its timing coincides with a possible wave-like structure in front of the CME, likely associated with compressed solar wind plasma. However,  the limitations in the observations of this wave-like structure make this a matter of speculation.

As the comet subsequently encounters the CME flank, the kink evolves and becomes clearly disconnected, approximately 6.5 hours later, consistent with the flank crossing of the CME as determined from 3D geometry. The disconnection and the detached tail are marked by a yellow arrow in panel (e). Note that by $\sim$18 UT on 2023 September 12, a front associated with a new CME appears in the FOV (panel g). After the disconnection, the tail progressively regrows (from panels f to g) as analyzed and discussed in the next Section~\ref{sec:regrowth}. 

\begin{figure}[ht]
\begin{center}
\includegraphics[width=6.6cm]{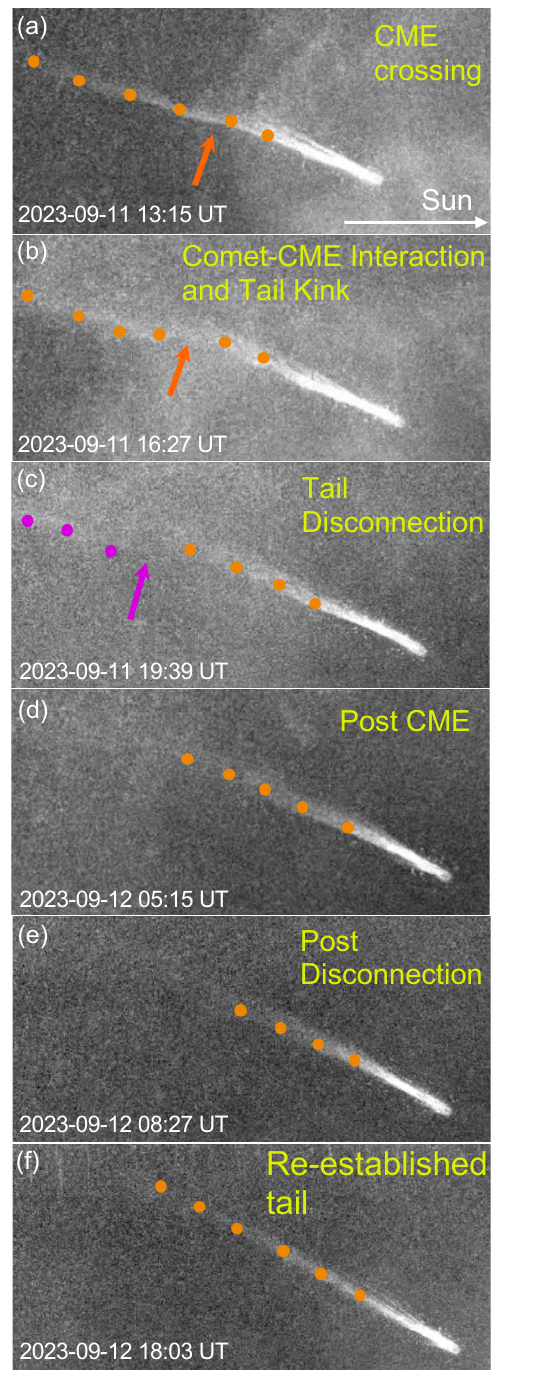}
\caption{Time sequence of selected SoloHI image frames illustrating the tail dynamics and disconnection event in a zoomed view of the comet in the subfield. The orange circles mark the intensity across the tail at a few selected regions along the tail, whereas the magenta circles indicate those for the disconnected tail. The white arrow in panel (a) indicates the sunward direction, toward the right side of the subfield.}\label{Figure3}
\end{center}
\end{figure}

\subsection{Characterization of the Tail Regrowth after Disconnection }\label{sec:regrowth}
A time sequence of SoloHI images tracing the close-up details of the comet’s tail to examine its morphology and the TDE dynamics is shown in Figure~\ref{Figure3}. Each panel in the figure is annotated with the corresponding stage of the disconnection. In panels (a) and (b), the tail appears continuous, marked by orange circles, with the kink discernible, and grows, as marked by the orange arrows. In panel (c), the kink disappears (magenta arrow), and a segment of the tail is seen disconnected from the nucleus (magenta circles), marking the onset of the disconnection. The downstream location of the disconnection along the tail, far from the nucleus, suggests that the magnetic interaction responsible for the disconnection is initiated where the CME flank's magnetic field intersects the pre-existing tail. The following two panels (d and e) show the evolution of the tail after the CME passage and the resulting tail disconnection. The last panel (f) captures the emergence of a new, straighter tail from the coma, which eventually grows back to its pre-event extent. 

\begin{figure*}[ht]
\begin{center}
\includegraphics[width=18cm]{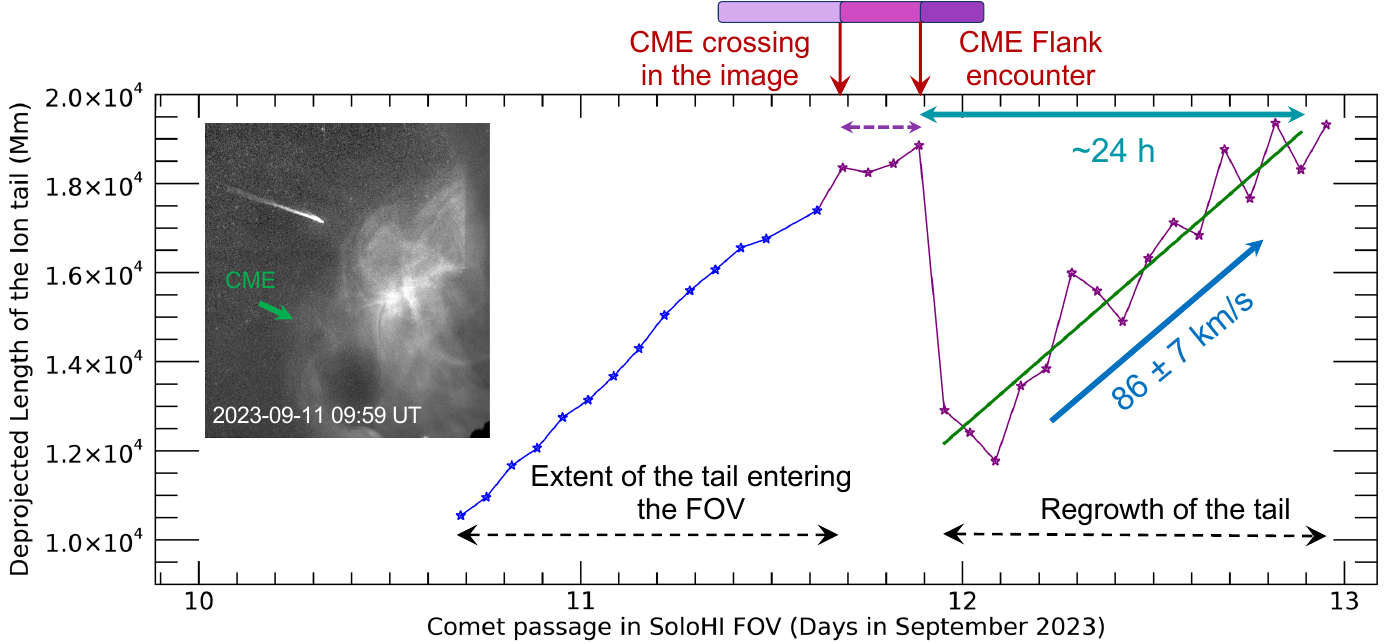}
\caption{Rate of regrowth of the ion tail after the disconnection event. The inset figure shows the comet and the CME as they propagate in the SoloHI tile 2 FOV. Abrupt decrease in tail length near 19:00 UT on 2023 September 11 indicates the tail disconnection, following the CME crossing and flank encounter (red arrows on the top). The extent of the comet's tail entering the FOV is shown in blue symbols in the plot. The tail remains fully within the FOV for four consecutive frames (marked by the purple arrow), after which the disconnection and regrowth are observed. The purple overbar marks the CME’s passage duration through the SoloHI FOV, CME's entry (light purple), the leading edge passage in projection before the comet-CME interaction (medium), and the interval after the comet crosses the CME flank (dark). The growth rate is determined from the slope of the linear fit indicated by the green line.}
\label{Figure4}
\end{center}
\end{figure*}
To quantify how the tail re-establishes after disconnection, we measure its length in each SoloHI image. A photometric intensity threshold ($\sim$ $1.8\times10^{-12}$ MSB; Mean Solar Brightness unit) slightly above the background intensity is used to determine the reliable outermost detectable extent of the tail in each image \footnote{The bright comet head intensity is well above $5.9\times10^{-12}$ MSB}. We acknowledge that the tail might extend beyond the measured length, and the apparent endpoint of the tail brightness falls to the lowest observable magnitude that the photometric sensitivity of the tail is capable of detecting. Even so, applying a consistent intensity threshold through time provides an objective measure of the return of the tail to the pre-event state. Moreover, in this examination of the regrowth rate, we emphasize tracking the tail's recovery back to the pre-event extent rather than the absolute magnitude of the tail length.

During our analysis period, the Sun-comet-Orbiter phase angle varies only minimally (from $\sim$43 to $38^\circ$), so viewing geometry effects are negligible, making this length comparable across frames. We measure the length of the tail as seen in tile 2 along the instantaneous tail axis in each frame, as the tail's direction can swing with the IMF. The projected length in pixels (i.e., the linear distance from the comet's nucleus to the farthest point visible along the tail axis) is converted to megameters (Mm) by considering the SoloHI tile 2 plate scale (angular size per pixel of $74\arcsec$) and the corresponding spacecraft-to-comet distance shown in Figure~\ref{Figure1}. Using the comet's 3D heliocentric position, the Solar Orbiter LOS geometry, and assuming a radial propagation direction of the tail away from the Sun, the measured tail lengths are deprojected to account for line-of-sight projection effects.

In Figure~\ref{Figure4}, we show the resulting measurements of the tail's length as the comet transits the tile 2 FOV for three days. The initial rise in tail length observed in Figure~\ref{Figure4} is not physical as it merely denotes the increase in the fraction of tail that enters tile 2 as the comet moves closer to the center of the FOV. Although the tail is also visible in tile 3 during this period, the difference in timing between the two tiles, as well as the distinct photometric sensitivity, would leave any attempt to measure the tail across the tile boundary prone to significant errors. Instead, we consider the plateau region observed across four consecutive timestamps, marked by the purple arrow in Figure~\ref{Figure4}, as representative of the full tail length in tile 2. 

The abrupt decrease in the tail length around 19:00 UT on 2023 September 11 indicates the tail disconnection. The CME evolution is shown by the purple overbar at the top of the plot, marking the interval during which the CME propagates in the SoloHI FOV and encounters the comet. The first red arrow from left indicates the time when the CME leading edge is seen crossing the comet in the SoloHI images in projection, while the second arrow marks the timing of the actual CME–comet flank encounter discussed in Section~\ref{sec:CMEkin}.

Between the TDE and the tail exiting the FOV, the length of the reforming tail increases approximately linearly with time. This linear increase gives the tail regrowth at a rate of $\sim$\(86\pm7\,\mathrm{km}\,\mathrm{s}^{-1}\). For context, the corresponding plane-of-sky regrowth rate is $\sim$\(53\pm5\,\mathrm{km}\,\mathrm{s}^{-1}\), which can be regarded as a lower limit. The recovery to the pre-event extent occurs over \(24\,\mathrm{h}\). This regrowth rate is well below the typical bulk slow solar wind speeds ($\sim$\(300 - 400\,\mathrm{km}\,\mathrm{s}^{-1}\)) \citep{Parker_1958, Brooks_2015} at those heliocentric distances. This rate reflects the rebuilding of the tail, which can be governed by processes such as the ion production rate of the comet, magnetic field reconfiguration, or the re-draping of the field as the post-CME environment relaxes, as discussed in \citet{Jockers_1973, Jones_2000, Voelzke_2005, Opitom_2016}. 

\subsection{Dynamics of the Detached Tail}\label{Dynamics_Section}

After the TDE, we track the motion of the drifting detached tail (magenta circles in Figure~\ref{Figure3}c) to assess its dynamics and to place its motion in the context of the CME. Using the comet–spacecraft viewing geometry as described in Section~\ref{sec:regrowth}, we measure the speed of the detached comet tail segment to determine its antisunward drift relative to the disconnection site. Assuming that the detached plasma segment moves approximately radially outward from the head of the comet, the displacement measured between successive frames gives thedeprojected drift speed of the trailing edge of $\sim295\pm20$ km\,s$^{-1}$. This speed is slightly lower than the local CME flank speed of $\sim$\(380\pm20\,\mathrm{km}\,\mathrm{s}^{-1}\) inferred from the GCS reconstruction discussed in Section~\ref{sec:CMEkin}, acknowledging the expected uncertainties associated with the GCS reconstruction and with the projection effects inherent in imaging structures at different locations in space along a single line of sight. 

This drift speed indicates that the detached material is probably transported by the flank of the CME as it sweeps across the comet after the interaction. This behavior is naturally explained by the comet’s large angular separation from the CME nose ($60^\circ$ at the time of TDE). Furthermore, the combination of this drift speed, along with the fluctuations observed in the comet tail during the initial wave-like structure and CME leading-edge crossing discussed in Figure~\ref{Figure2}, and the 6.5-hour delay to the onset of disconnection, further points to a trigger arising within the CME's magnetic structure at the flank. 

\section{Conclusions}\label{sec:conclusion}
SoloHI has provided images of multiple TDEs of the C/2023 P1 comet during its passage through the inner heliosphere between 2023 September 01 and 14, capturing the comet at distances of 0.55 to 0.26 au from the Sun. In this study, we analyzed the TDE that occurred on 2023 September 11, six days before the comet's perihelion. In the following, we summarize the key findings and their broader implications. 

\begin{itemize}
  
  \item Contrary to the prevailing expectations that HCS crossings, SIRs, and CMEs drive TDEs, we find that all four events observed here are associated with CME crossings. The SoloHI image sequence shows that the CME passage produces sustained fluctuations, large-amplitude kinks in the ion tail over a span of a few hours, and ultimately produces a tail disconnection of a portion of the pre-existing tail of the comet. The combined analysis on the timing, 3D geometry, and observed tail dynamics demonstrates that the trigger arises from the passage of the CME’s flank, providing a clear example of a magnetically driven, CME-induced TDE in the inner heliosphere.

  \item \textit{Regrowth Rate:} Following the disconnection, the ion tail returns to its pre-event length ($1.9\times10^6$ km) over $\sim$\(24\,\mathrm{h}\), yielding a linear growth rate of $\sim$\(86\pm7\,\mathrm{km}\,\mathrm{s}^{-1}\) ($\sim$\(10.7\pm0.6\,\mathrm{R_{\odot}\,{day}^{-1}}\)). Previous space-based studies of cometary TDEs, \cite[e.g.,][]{Vourlidas_2007, Kuchar_2008} established that CMEs can lead to disconnection of comet tails and provided diagnostics of the tail and disconnection dynamics. This study represents the first quantitative measurement of a cometary tail regrowth rate. This measurement is made possible by the unique SoloHI observations with sufficient photometric sensitivity, exposure time, and benefited by a fortuitous viewing geometry that resolved the cometary plasma dynamics and tail disconnections with the comet's path through the FOV at near-Sun heliocentric distances $\leq$0.32 au. While several physical processes may influence the rebuilding of the tail, a detailed interpretation of these mechanisms with the implications for cometary science is beyond the scope of this work.

  \item \textit{Disconnection Geometry:} The TDE occurs $\sim$6.5 hours after the CME's leading edge first appears, by projection, to pass over the comet in SoloHI images. The pronounced kink in the tail observed beforehand may reflect the comet's interaction with a CME-driven disturbance ahead of the flank, which distorts the tail but is insufficient to sever it. The actual disconnection takes place only when the CME's flank magnetic structure reaches the comet, consistent with the comet’s $\sim$$60^\circ$ angular separation from the CME nose and its location within the broad flank of the CME flux rope.
  
  After the tail disconnects, the detached tail segment drifts antisunward at a deprojected speed of $\sim$\(295\pm20\,\mathrm{km\,s^{-1}}\), comparable to the local flank speed at the location where the comet interacts with the CME. This drift speed indicates that the tail is being passively transported outward by the slowly expanding CME’s flank region. This dynamic behavior, together with the geometric offset from the CME direction, reinforces that the comet experienced a flank–driven interaction rather than a direct nose encounter.
  
\end{itemize}

This analysis establishes a benchmark for understanding the physics of tail reformation in the inner heliosphere and provides observational constraints relevant to ion production, plasma pickup, and magnetic draping. Collectively, the measured regrowth behavior, disconnection timing, and drift speed show that cometary ion tails respond sensitively to the magnetic structure of CMEs during comet–CME interactions. Likewise, the several-hour delay on the disconnection onset and regrowth timescale highlights the ability to resolve and examine the highly dynamic comet’s near-nucleus plasma environment at close heliocentric distances. These observations demonstrate, for the first time, that tail reformation is a gradual, quantifiable process responding to the evolving ambient environment rather than appearing instantaneous on observational timescales.


\begin{acknowledgments}
Solar Orbiter is a mission of international cooperation between the European Space Agency (ESA) and the National Aeronautics and Space Administration (NASA), operated by ESA. The Solar Orbiter Heliospheric Imager (SoloHI) instrument was designed, built, and is now operated by the US Naval Research Laboratory with support from the NASA Heliophysics Division, Solar Orbiter Collaboration Office under DPR NNG09EK11I. The NRL effort was also supported by the Office of Naval Research. S.B.S. acknowledges the support from George Mason University (GMU) via the NRL contract (N00173-23-2-C603). GS and AV acknowledge funding support from the SoloHI Grant 80NSSC22K1028.
\end{acknowledgments}

\bibliographystyle{aasjournal} 
\bibliography{Comet_biblio}
\end{document}